# A BLOCK BASED SCHEME FOR ENHANCING LOW LUMINATED IMAGES

A.Saradha Devi[1], S. Suja Priyadharsini[2], S. Athinarayanan[3]

[1]ME-Applied Electronics Student, Anna University, Tirunelveli.
E-mail:saradhad054@gmail.com
[2]Lecturer, Department of ECE, Anna University,Tirunelveli.
E-mail:sujapriya_moni@yahoo.co.in.
[3] Team Leader, R&D Image Processing, Manatec electronics
E-mail:s_athi1983@yahoo.co.in.

## ABSTRACT

*In this paper the background detection in images in poor lighting can be done by the use of morphological filters. Lately contrast image enhancement technique is used to detect the background in image which uses Weber's Law. The proposed technique is more effective one in which the background detection in image can be done in color images. The given image obtained in this method is very effective one. More enhancement can be obtained while comparing the results. In this technique compressed domain enhancement has been used for better result.*

## 1. INTRODUCTION

In this Paper the concept is to detect the background in images in poor lighting. The contrast enhancement problem can be approached from various methodologies, among which is mathematical morphology(MM)[1]-[2]. For background detection three methods has been carried out. Other works based on the contrast mapping concept have been developed elsewhere [3]-[5]. Even though morphological contrast has been largely studied, there are no methodologies. There are techniques based on data statistical analysis, such as global and histogram equalization. During the histogram equalization process, the grey level intensities are recorded within the image to obtain an uniform distributed histogram[6].

In the First method the background images in poor lighting of grey level images can identified by the use of morphological operators[7]. Lately image enhancement has been carried out by the application based on Weber's Law. After that erosion dilation and opening by reconstruction method is followed. Then the proposed method in which it is used to detect the background images in color images has been carried out.

In the First method the images is divided into 'n' no of blocks and each block is analyzed. Due to the problem called contouring effect occur in this method the next method called erosion dilation method is followed in that method also there are certain disadvantage over it. Hence a proposed method called opening by reconstruction method has been carried out. Lately DCT domain technique is used to process the color images.

The Dynamic range of Intensity values may be small due to the presence of strong background illumination. This problem gets more complicated when the illumination of the scene widely varies in the space. In such case it necessary to the local contrast of the image. Image enhancement often deals with such improvement of the image contrast.

A Majority of techniques advanced so far have focused on the enhancement of grey images in the spatial domain. Some of the techniques like adaptive histogram equalization, constant variance enhancement, homomorphic filtering, high-pass filtering, etc.. can be used for image enhancement technique[8] but all these methods are based on spatial domain techniques only. It is observed that by the use of compressed domain technique only we can able to get better enhancement, for that only we are moving for DCT domain technique.

The Discrete Cosine Transform (DCT) is a technique that converts a spatial domain technique waveform into its constituent frequency components as represented by a set of coefficients. The process of reconstructing a set of spatial domain samples is called the Inverse Discrete Cosine Transform(IDCT). 2D-DCT has most often implemented by employing row-column or column-row decomposition and operating 1-D DCT on row and column data separately. This DCT transformation consist of all coefficients with it. The coefficients are like AC,DC.

AC coefficients of the DCT image block represents less energy, however they can be estimated from DC coefficients of neighboring DCT. AC coefficients are required to reduce the blocking artifacts. The main concept in the color image processing is the conversion of the R-G-B to $Y$-$C_b$-$C_r$ coordinates. Why the conversion process is necessary means, sometimes if we enhance the image without conversion process means there may be a chance of getting wrong output image.

### 1.1. Mathematical Morphology

Mathematical morphology (MM) is a theory and technique for the analysis[1]and processing of geometrical structures, based on set theory, lattice theory, topology, and random functions. MM is most commonly applied to digital images, but it can be employed as well on graphs, surface meshes, solids, and many other spatial structures. The morphological operator is a composition of three basic operators: a dilation, an erosion of the input image by the input structuring element and a subtraction of these two results. As erosions and dilations, the key mechanism under the opening operator is the local comparison of a shape, the structural element, with the object that will be transformed. If, when positioned at a given point, the structural element is included in the object than the whole structural element will appear in the result of the transformation, otherwise none of its points will appear. One important application of the morphologic gradient in binary images is to find their boundaries.

The same operator applied to gray level images, but without filtering it does not produce good results because it is very sensible to noise.

## 2. Image Enhancement and Contrast Enhancement

The goal of Image enhancement include the improvement of the visibility and perceptibility of the various regions into which an image can be partitioned and of the detect ability of the image features inside the regions. These goals include tasks such as cleaning the image from various types of noise enhancing the contrast among adjacent regions or features, simplifying the image via selective smoothing or elimination of features at certain scales and retaining only features at certain desirable scales. Image enhancement is usually followed by (or is done simultaneously with) detection of features such as edges, peaks, and other geometric features which is of paramount importance in low-level vision. Further, many related vision problems involve the detection of a known template, such problems are usually solved via template matching.

Imagine a gray level image that has resulted from blurring an original image by linearly convolving it with a Gaussian function of variance. This Gaussian blurring can be modeled by running the classic heat diffusion differential equation for the time interval starting from the initial condition at t=0. If we can reverse in time this diffusion process, then we can de-blur and sharpen the blurred image. By approximating the spatio-temporal derivatives of the heat equation with differences, we can derive a linear discrete filter that can enhance the contrast of the blurred image by subtracting from a discretized version of its Laplacian This is a simple linear deblurring scheme, called unsharp contrast enhancement.

### 2.1. Morphological Transformations and Weber's Law

In mathematical morphology, increasing and idempotent transformations are frequently used. Morphological transformations complying with these properties are known as morphological filters[1] the basic morphological filters are the morphological opening and morphological closing using a structural element. The morphological opening and closing can be expressed as follows.

The International journal of Multimedia & Its Applications (IJMA) Vol.2, No.3, August 2010

$$\gamma_{\mu B}(f)(x) = \delta_{\mu B}(\varepsilon_{\mu B}(f))(x)$$
$$\varphi_{\mu B}(f)(x) = \varepsilon_{\mu B}(\delta_{\mu B}(f))(x) \qquad (1)$$

On the other hand we will use the size of the structuring element size 1 or any size µ. Size 1 means a square of 3X3 pixels. While size µ means a square of ( 2µ+1), ( 2µ+1) size.

Weber's Law states that the it is the ratio of the difference in max to min Luminance value to the min Luminance value and it is denoted by C

$$C = \frac{L_{max} - L_{min}}{L_{min}} \qquad (2)$$

## 2.2. Opening and Closing

A study of a class of openings and closings is investigated using reconstruction criteria. The main goal in studying these transformations consists of eliminating some inconveniences of the morphological opening (closing) and the opening (closing) by reconstruction. The idea in building these new openings and closings comes from the notions of filters by reconstruction and levelings. In particular, concerning the notion of levelings, a study of a class of lower and upper levelings is carried out.

The original work of levelings is due to Meyer, who proposes this notion and introduces some criteria to build the levelings in the general case (extended levelings and self-dual transformations). We see the criteria proposed by Meyer as reconstruction criteria during the reconstruction process from a marker image into the reference image. We show that new openings and closings are obtained, enabling intermediate results between the traditional opening (closing) and the opening (closing) by reconstruction. Some applications are studied to validate these transformations.

## 2.3. Image Background Analysis by Blocks

When D represents the digital space under study, with $D=Z^2$ and Z the integer set. In this way , let D be the domain of definition of the function f. The image f is divided into n blocks $w^i$ of size $l_1 X l_2$ Each block is a sub image of the original image. The maxima and minima intensity values in each subimage is denoted as $M_i, m_i$

For each analyzed block the background criteria $\tau_i$ is found out by the following way.

$$\tau_i = \frac{m_i + M_i}{2} \quad \forall i = 1,2,....,n. \qquad (3)$$

Where $\tau_i$ represents a division line between the clear (f> $\tau_i$) and dark (f< $\tau_i$) intensity level. The grey level used in this is a constant one and which is given to be

$$k_i = \frac{255 - m_i^*}{\log(256)} \quad \forall i = 1,2,....,n. \qquad (4)$$

51

The International journal of Multimedia & Its Applications (IJMA) Vol.2, No.3, August 2010

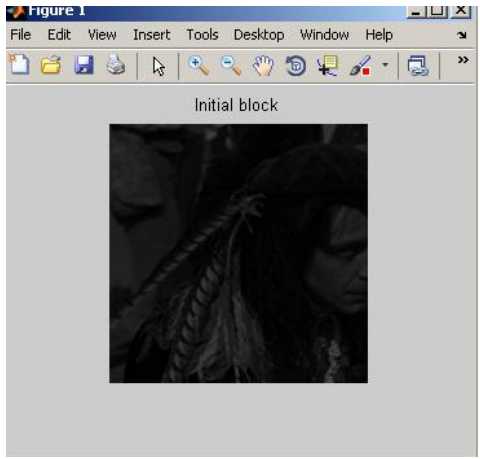 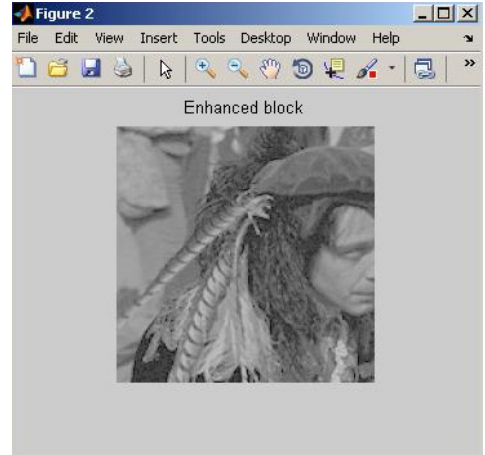

Figure:1 a)Initial Block          b)Enhanced Block

## 2.4. Image Background Analyzed By Opening By Reconstruction

Instead of dividing the original images into no of blocks and without using the Erosion and Dilation property a new method is used here. In this method the morphological transformations generate a new contour when the structuring element is increased. While by using morphological erosion or dilation which touches the regional min and merges it with the regional maxima to detect the background criteria. In this method the background detection method is same as the above method but the only thing is the way to detect the background is modified and the background criteria detection is given by the following expression

$$\tau(x) = \tilde{\gamma}_\mu(f)(x). \qquad (5)$$

By having the value of the structuring element size to be a constant one it is observed that we can able to get a clear image.

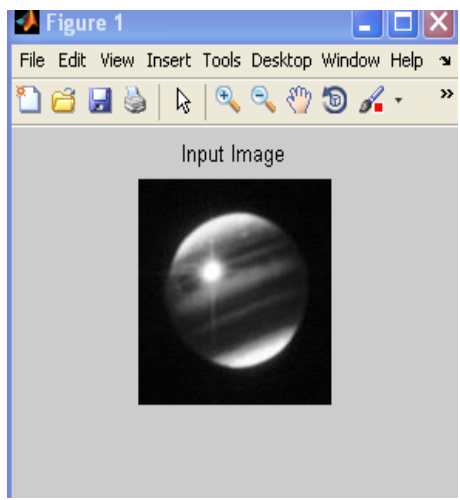 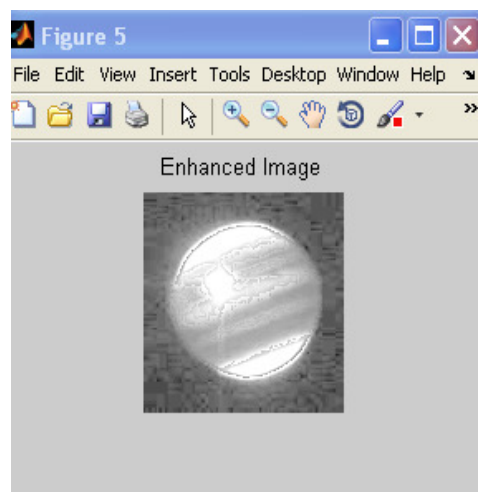

Figure:2 a)Input Image          b)Enhanced Image

## 2.5. Multi Background Image Analysis

By the use of image background detection using opening by reconstruction method we obtain a clear image comparing to the Dilation and Erosion method.

If we have a constant erosion and dilation value means the amount of new contours generation will be in lesser value only. But it is observed that by varying the value of the structuring element size we can able to get clearer image than by having a constant structuring element size.

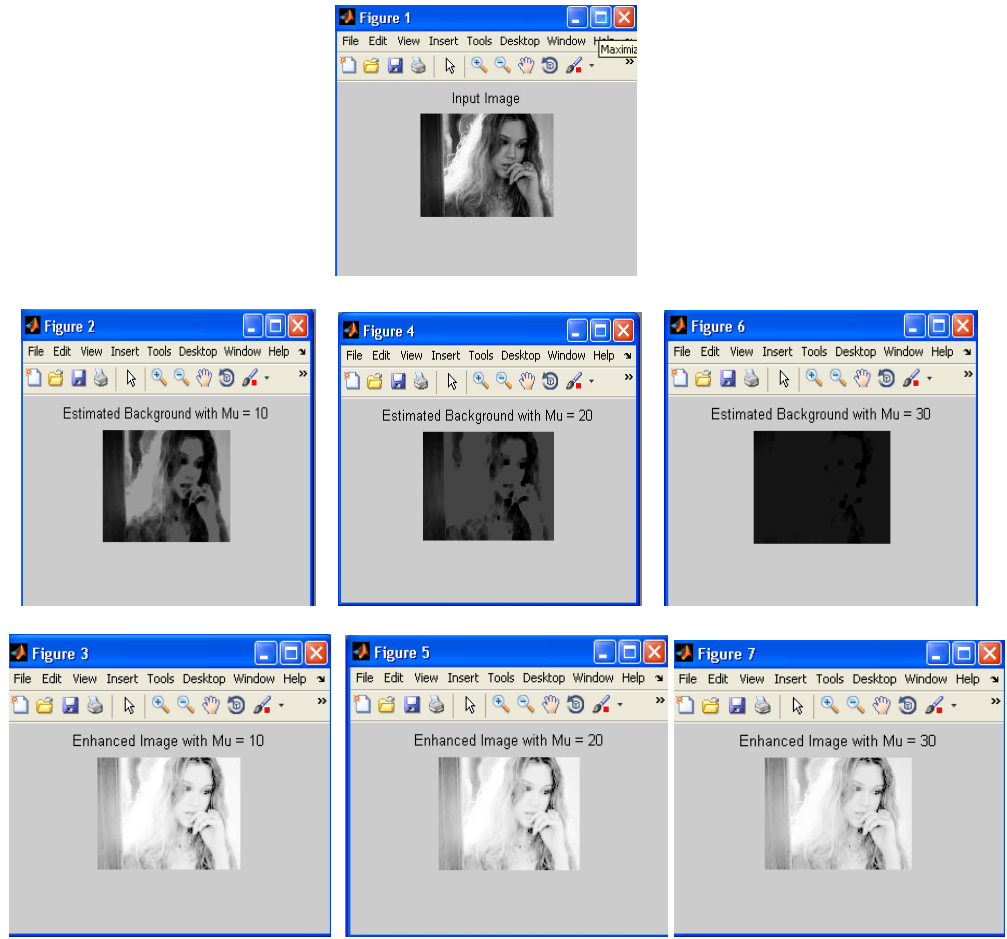

Figure:3 a)Input Image b)Estimated Background C)Enhanced Image

## 2.6. Performance Metrics

We have compared the opening by reconstruction method with the dilation erosion method by measuring the parameters SSIM values and Normalized Entropy.

### 2.6.1. SSIM Values

The structural similarity (SSIM) index is a method for measuring the similarity between two images. The SSIM index is a full reference metric, in other words, the measuring of image quality based on an initial uncompressed or distortion-free image as reference. SSIM is designed to improve on traditional methods like peak signal-to-noise ratio (PSNR) and mean squared error (MSE), which have proved to be inconsistent with human eye perception. The SSIM metric is calculated on various windows of an image. The measure between two windows of size *NXN x* and *y* is:

$$SSIM(x, y) = \frac{(2\mu_x\mu_y + C_1)(2\sigma_{xy} + C_2)}{(\mu_x^2 + \mu_y^2 + C_1)(\sigma_x^2 + \sigma_y^2 + C_2)} \quad (6)$$

With $\mu_x$ the average of x, $\mu_y$ the average of y, $\sigma_x^2$ the variance of x, $\sigma_y^2$ the variance of y, $\text{cov}_{xy}$ the covariance of x and y, $C_1 = (K_1 L)^2$, $C_2 = (K_2 L)^2$ two variables to stabilize the division with weak denominator; $L$ the dynamic range of the pixel-values, $k_1 = 0.01$ and $k_2 = 0.03$ by default.

SSIM values

| Input Image | Dilation Erosion Method | Opening by Reconstruction Method |
|---|---|---|
| Image1 | 0.16165 | 0.16936 |
| Image2 | 0.18291 | 0.25859 |
| Image3 | 0.16446 | 0.16626 |
| Image4 | 0.27362 | 0.30875 |
| Image5 | 0.27362 | 0.30875 |
| Image6 | 0.26757 | 0.29735 |
| Image7 | 0.20424 | 0.21331 |

### *2.7. Normalized Entropy*

Shannon entropy is a measure of the information contained in the image and it defined as,

$$H = -\sum_{x=0}^{N_{GL}-1} p(x) \log_2(p(x)) \quad (7)$$

Where p(x) is the probability of GL x occurring in the image. In order to compare the processing effect on a population of images with widely divergent H values, normalized entropy, $H_N$, was adopted. This is defined as the ratio of the entropy of the treated image to that of the original image.

Normalized Entropy

| Input Image | Dilation Erosion Method | Opening by Reconstruction Method |
|---|---|---|
| Image1 | 0.8947 | 0.9557 |
| Image2 | 0.8641 | 0.9394 |
| Image3 | 0.9160 | 0.9540 |
| Image4 | 0.8904 | 0.9627 |
| Image5 | 0.8942 | 0.9209 |
| Image6 | 0.9137 | 0.9384 |
| Image7 | 0.8847 | 0.9079 |

## 3. DCT

The Discrete Cosine Transform (DCT) is a fourier related transform similar to the Discrete Fourier Transform (DFT). But using only real numbers. It is equivalent to a DFT of roughly twice the length, operating on real data with even symmetry, where in some variants the input and/or output data are shifted by half a sample. The most common variant of discrete cosine transform is the type-II DCT, which is often called simply "the DCT"; its inverse, the type-III DCT, is correspondingly often called simply "the inverse DCT" or "the IDCT".

Formally, the discrete cosine transform is a linear, invertible function $F : \mathbf{R}n \rightarrow \mathbf{R}n$ (where $\mathbf{R}$ denotes the set of real numbers), or equivalently an $n \times n$ square matrix. There are several variants of the DCT with slightly modified definitions. The $n$ real numbers $x0, ..., xn-1$ are transformed into the $n$ real numbers $f0, ..., fn-1$ according to one of the formulas:

The inverse of DCT-I is DCT-I multiplied by $2/(n-1)$. The inverse of DCT-IV is DCT-IV multiplied by $2/n$. The inverse of DCT-II is DCT-III multiplied by $2/n$ (and vice versa). Like for the DFT, the normalization factor in front of these transform definitions is merely a convention and differs between treatments. For example, some authors multiply the transforms by $\sqrt{2/n}$ so that the inverse does not require any additional multiplicative factor.

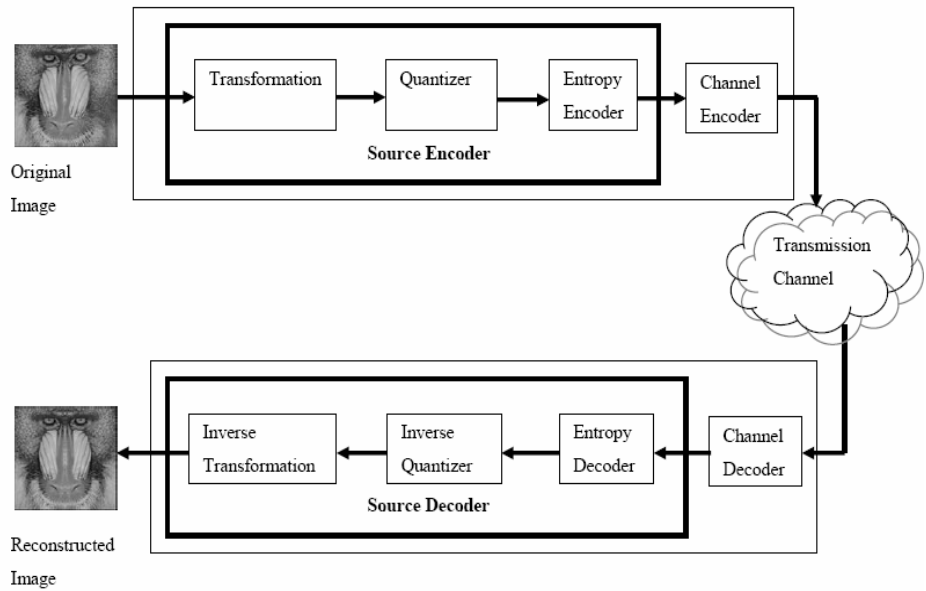

Figure:4 Basic block diagram for DCT conversion

The DCT is the core transform of many image processing application for reduced bandwidth. For image background detection in color images this DCT Scaling is used. Like other transformation DCT attempts to decorrelate the image data. After decorrelation each transformation coefficient can be encoded independently without losing compression efficiency.

Several algorithm and architecture has been proposed to optimize DCT scaling used 1-D and 2-D Algebraic integer quantization(AIQ).In this DCT scaling is used for better enhancement. Different algorithms has been carried out for DCT domain block they are such as Alpha Routing and Multicontrast enhancement.

Mainly the image block is compressed by the use of compressed domain technique in order to reduce the computational complexity.

### 3.1. Properties of DCT

The main properties of DCT are
- Decorrelation
- Energy Compaction
- Seperability
- Symmetry
- Orthogonality

## 3.2. Color Image Enhancement

The main objective of the image enhancement[8] is to modify attributes of an image to make it more suitable for a given task and specific observer. Based on image data representation space, image enhancement is divided into spatial domain and compressed domain. The process of converting from one form to another is called as transcoding.

### *3.2.1. Spatial Domain and Compressed Domain Techniques*

Image Processing in spatial domain can be expressed by

$$g(m,n) = T(f(m,n))$$

Where f(m,n) is the input image, and g(m,n) is the processed image and T is the operator defining the modification process.

The operator T is a monotone function and that can operate on individual pixels or a region in an image.

Typically color information is measured in compressed domain using DC image sequence and represented by DC color histograms. AC coefficients in the 8X8 DCT blocks are classified into frequency bands that roughly corresponds to smooth areas, horizontal and vertical edges. DCT blocks corresponding to the smooth region have the low frequency and have only very few AC coefficients and non-zero and only the DC coefficient. Once the original image is converted and splitted into N no of block means which consist of AC and DC coefficients of the original image. The image is processed in a zigzag manner. The first point present in the matrix form gives the information of the DC coefficient and all other things represents the AC coefficient.

## 3.3. Proposed Method

First the input color image obtained is converted into other format like HSV or $Y,C_bC_r$ formats because the color image obtained cannot be processed directly. Mostly $YC_bC_r$. format is preferred than the HSV conversion due to its less complexity. In $YC_bC_r$. the Y component represents the brightness of a pixel, the Cb and Cr components represent the chrominance (split into blue and red components). Y′CbCr is not an absolute color space, it is a way of encoding RGB information. The actual color displayed depends on the actual RGB colorants used to display the signal. Therefore a value expressed as Y′CbCr is only predictable.

The YCbCr color space conversion allows greater compression without a significant effect on perceptual image quality (or greater perceptual image quality for the same compression). The compression is more efficient because the brightness information, which is more important to the eventual perceptual quality of the image, is confined to a single channel, more closely representing the human visual system.

This conversion to YCbCr is specified in the JFIF standard, and should be performed for the resulting JPEG file to have maximum compatibility. However, some JPEG implementations in "highest quality" mode do not apply this step and instead keep the color information in the RGB color model, where the image is stored in separate channels for red, green and blue luminance. This results in less efficient compression, and would not likely be used if file size were an issue.

Once the image is converted then it is divided into NXN blocks and in that each block[13] consist of both the AC and DC coefficients of the image. The first component in the NXN matrix represents the DC value of the image and all other things represents the AC value of the image.

Now we are going to change the values of AC as well as DC coefficients in order to improve the brightness and contrast and color values of the color image. By the use of this method the background in color image can be obtained.

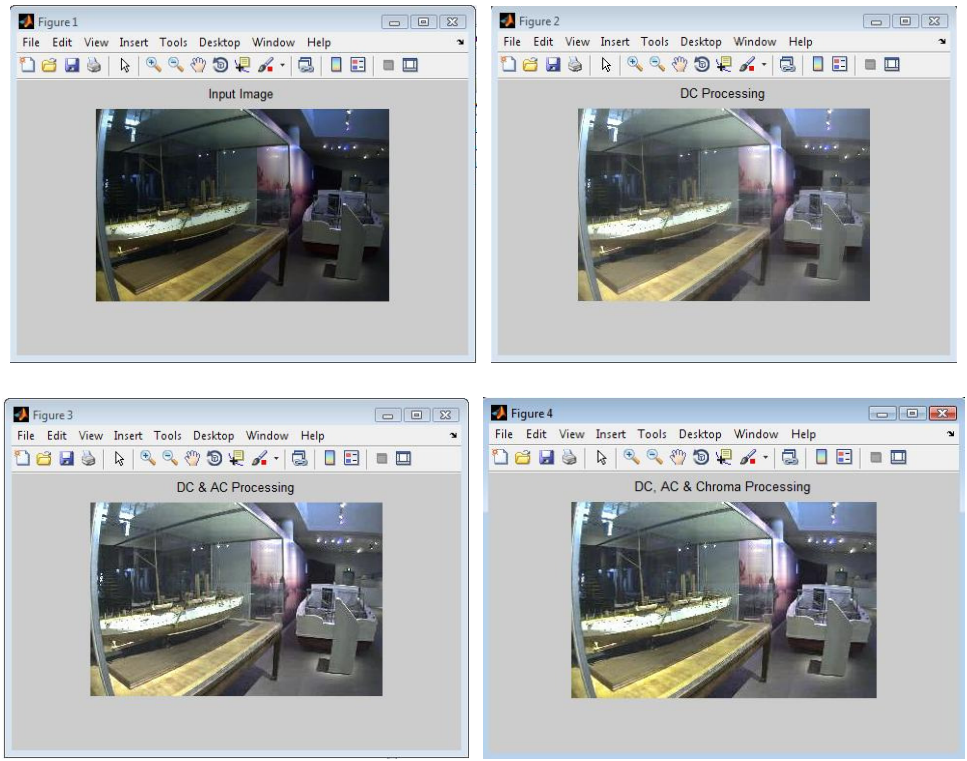

Figure:5 a)Input Image  b)DC processing  c)DC and AC processing  d)DC and AC with Chrominance processing

### 3.3.1. Methods of Comparision

- DC value is changed by keeping AC value as constant
- AC value is changed by keeping DC value as constant
- Both AC and DC values are changed.

These are the methods we are going to compare ie the output images obtained is compared to get better enhancement.

By keeping the DC value as constant and changing the AC value only will give a Black output image.

By keeping the AC value as constant and changing the DC value only will give a White output image. In both these process no color information will be present in the image.

By changing both the AC and DC coefficients of the image we can able to get better results of image.

## 3.4. Performance Matrices

We compared the three mapping functions for several images by measuring the parameters JPQM metric and CEF.

**Image-1**

| Techniques | DC Processing | | DC and AC Processing | | DC,AC with Chrominance Processing | |
|---|---|---|---|---|---|---|
| | JPQM Metric | CEF | JPQM Metric | CEF | JPQM Metric | CEF |
| Twisting Function | 6.9318 | 0.8884 | 7.5642 | 0.8809 | 7.5423 | 1.1305 |
| Eta Function | 8.3248 | 0.9997 | 7.9459 | 0.9992 | 7.9430 | 0.9365 |
| S Function | 7.9779 | 0.9930 | 7.8670 | 0.9907 | 7.8601 | 1.0311 |

**Image-2**

| Techniques | DC Processing | | DC & AC Processing | | DC,AC with Chrominance Processing | |
|---|---|---|---|---|---|---|
| | JPQM Metric | CEF | JPQM Metric | CEF | JPQM Metric | CEF |
| Twisting Function | 6.8672 | 0.9773 | 8.1107 | 0.9650 | 8.0889 | 1.2506 |
| Eta Function | 8.9549 | 0.9997 | 8.3451 | 0.9994 | 8.3430 | 0.9363 |
| S Function | 8.4299 | 0.9981 | 8.1647 | 0.9964 | 8.1618 | 1.0377 |

**Image-3**

| Techniques | DC Processing | | DC & AC Processing | | DC,AC with Chrominance Processing | |
|---|---|---|---|---|---|---|
| | JPQM Metric | CEF | JPQM Metric | CEF | JPQM Metric | CEF |
| Twisting Function | 7.4415 | 0.9366 | 8.2780 | 0.9134 | 8.2656 | 1.0408 |
| Eta Function | 9.0994 | 1.0006 | 8.5968 | 1.004 | 8.5966 | 0.9454 |
| S Function | 8.6036 | 0.9985 | 8.3267 | 0.9968 | 8.3260 | 1.0065 |

**Image-4**

| Techniques | DC Processing | | DC & AC Processing | | DC,AC with Chrominance Processing | |
|---|---|---|---|---|---|---|
| | **JPQM Metric** | **CEF** | **JPQM Metric** | **CEF** | **JPQM Metric** | **CEF** |
| **Twisting Function** | 7.4684 | 0.9463 | 8.0652 | 0.9351 | 8.0595 | 1.1209 |
| **Eta Function** | 8.7214 | 0.9980 | 8.3911 | 0.9977 | 8.3990 | 0.9169 |
| **S Function** | 8.2178 | 0.9973 | 8.0189 | 0.9958 | 8.0170 | 1.0261 |

**Image-5**

| Techniques | DC Processing | | DC & AC Processing | | DC,AC with Chrominance Processing | |
|---|---|---|---|---|---|---|
| | **JPQM Metric** | **CEF** | **JPQM Metric** | **CEF** | **JPQM Metric** | **CEF** |
| **Twisting Function** | 8.0813 | 0.9026 | 9.1264 | 0.8678 | 9.1171 | 1.0967 |
| **Eta Function** | 9.4687 | 0.9997 | 9.1452 | 0.9948 | 9.1288 | 0.9180 |
| **S Function** | 9.3254 | 0.9948 | 9.1765 | 0.9854 | 9.1772 | 1.0240 |

## Conclusion

To detect the background images and to enhance the contrast in grey level with poor lighting, a methodology was introduced, however a difficulty was detected by this method hence several analysis has been done, Morphological contrast enhancement transformations has introduced. These contrast transformations are characterized by the normalization of grey level intensities, and it is observed that by the use of the opening by reconstruction or by the use of erosion dilation methods we can only able to adjust the brightness and contrast of the images and we cannot able to adjust the color information present in the image. In order to process the color information present in an image can be done by the proposed method, which is used to adjust the color information present in the original image, which is more better than the previous methods. By the use of the DCT technique both the AC & DC coefficients are adjusted separately, Here we are enhancing the color images in the block by DCT domain. In this the chromatic component in addition to the process of luminance component gives better visualization on images. The time constraints are also get reduced by the use of this method. A comparative study with different other schemes can also be carried out here. It is found that the proposed scheme output performs well comparing to other methods.

**Authors**

**A.Saradha Devi** received her B.E degree from Anna University, Chennai, India. Currently, she is pursuing M.E in Anna University, Tirunelveli, Tamilnadu, India. She published two research papers in national conferences. Her main research interest includes image processing.

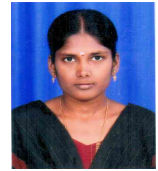

**S.Suja Priyadharsini** received her B.E degree from Manonmanian Sundaranar University, Tirunelveli. She received her M.E degree in Applied Electronics from Anna University, Chennai, India. She is pursuing her Ph.D from Anna University, Tirunelveli, Tamilnadu, India. She is currently working as a Lecturer in ECE Department in Anna University, Tirunelveli. Her main research interest includes soft computing, signal Processing and medical electronics.

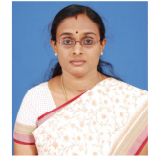

**S.Athinarayanan** received her B.E degree from Anna University, Chennai, India. He published several research papers in national conferences and journals. His main research interest includes image processing and computer vision.

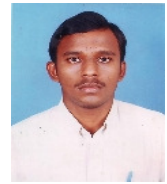